# Infinite potential well with a sinusoidal bottom

A. D. Alhaidari[a,b,*] and H. Bahlouli[b]

[a] *Shura Council, Riyadh 11212, Saudi Arabia*
[b] *Physics Department, King Fahd University of Petroleum & Minerals, Dhahran 31261, Saudi Arabia*

We construct a tridiagonal matrix representation of the wave operator that maps the wave equation into a three-term recursion relation for the expansion coefficients of the wavefunction. Finding a solution of the recursion relation is equivalent to solving the original problem. Consequently, a larger class of solvable potentials is obtained. The usual diagonal representation constraint results in a reduction to the conventional class of solvable potentials. To exhibit the power of this approach, we give an exact solution for the infinite potential well with sinusoidal bottom.

Exactly solvable potentials in nonrelativistic quantum mechanics have more than a pedagogical value and are not just for academic interest. The associated eigenvalues and eigenfunctions give insight into the physical concepts of quantum mechanics. Besides, some of these potentials could be used as the unperturbed part of a more realistic Hamiltonian. The search for exactly solvable potentials has been a subject of intense research since the advent of quantum mechanics. The factorization method [1], group theoretical techniques [2], supersymmetric quantum mechanics [3], and shape invariance [4] are few among many approaches used to find exact solutions of the wave equation. In these developments, the main objective is to find solutions of the eigenvalue wave equation $H|\psi\rangle = E|\psi\rangle$, where $H$ is the Hamiltonian and $E$ is the corresponding energy. Conventionally, one looks for a representation in which the Hamiltonian has a diagonal structure exhibiting the eigenvalues or the energy spectrum of the associated potential.

In this letter, we search for an algebraic representation of the eigenvalue problem in the space of square integrable discrete basis with elements $\{\phi_n\}_{n=0}^{\infty}$. The main contribution of our work results from relaxing the usual restriction of a diagonal matrix representation of the energy eigenvalue equation. We only require that it be tridiagonal and symmetric. That is, the action of the wave operator on the elements of the basis is allowed to take the general form $(H-E)|\phi_n\rangle \sim |\phi_n\rangle + |\phi_{n-1}\rangle + |\phi_{n+1}\rangle$. For example, in an orthogonal basis, one obtains

$$\langle\phi_n|H-E|\phi_m\rangle = (a_n - E)\delta_{n,m} + b_n\delta_{n,m-1} + b_{n-1}\delta_{n,m+1}, \qquad (1)$$

where the real coefficients $\{a_n, b_n\}_{n=0}^{\infty}$ are, in general, functions of the physical parameters of the problem. The hope is that by relaxing the diagonal constraint on the matrix representation of the wave operator, its solution space becomes large enough to incorporate new solvable potentials [5]. Moreover, this approach embodies powerful tools in the analysis of solutions of the wave equation by exploiting the intimate connection and interplay between tridiagonal matrix representations and the theory of orthogonal polynomials. In such analysis, one is at liberty to employ a wide range of well established methods and numerical techniques associated with these settings such as quadrature approximation and continued fractions [6].

---

[*] Corresponding Author, email: haidari@mailaps.org

The matrix wave equation, which is obtained by expanding $|\psi\rangle$ as $\sum_m f_m |\phi_m\rangle$ in $(H-E)|\psi\rangle = 0$ and projecting to the left on $\langle\phi_n|$, results in the following three-term recursion relation

$$E f_n = a_n f_n + b_{n-1} f_{n-1} + b_n f_{n+1}. \tag{2}$$

Consequently, the problem translates into finding solutions of the recursion relation for the expansion coefficients of the wave function, $\psi$. In most cases this recursion is solved easily and directly by correspondence with those of well known orthogonal polynomials. Equation (1) clearly shows that the usual diagonal representation is easily obtained by the requirement

$$b_n = 0, \text{ and } a_n - E = 0, \tag{3}$$

for all $n$, giving the discrete energy spectrum. However, if the physical parameters cannot satisfy this diagonal constraint, then the energy spectrum is obtained as discrete values of $E$ satisfying (2) when written as a matrix eigenvalue equation $E|f\rangle = H|f\rangle$. Next, we start by introducing the theoretical formulation of the problem and explain our approach. Thereafter, we apply it to obtain a solution for the special case of a potential well with a sinusoidal bottom, which has no known exact solution.

The one-dimensional time-independent Schrödinger equation for a point particle of mass $m$ in the field of a potential $V(x)$ is

$$\left[-\frac{\hbar^2}{2m}\frac{d^2}{dx^2} + V(x) - E\right]\psi(x,E) = 0. \tag{4}$$

The configuration space coordinate belongs to the physical interval $x \in [x_-, x_+]$, which could be finite, infinite, or semi-infinite. We make a transformation, $y = y(\lambda x)$, to a "reference" configuration space with coordinate $y \in [-1,+1]$, where $\lambda$ is a positive scale parameter having the dimension of inverse length. This transformation maps the wave equation (4) into

$$\left[(y')^2 \frac{d^2}{dy^2} + y'' \frac{d}{dy} + U(y)\right]\psi(y,E) = 0, \tag{5}$$

where $U(y(x)) = \frac{2m}{\lambda^2 \hbar^2}[E - V(x)]$ and the prime stands for the derivative with respect to $\lambda x$. Our approach is an algebraic one and thus we search for a complete discrete basis set that carries a faithful description of the wavefunction $\psi(y,E)$. Such a basis, has elements with the following general form: $\phi_n(y) = W(y) P_n(y)$, where $W(y)$ is a weight function and $P_n(y)$ is a polynomial of degree $n$ in $y$ and $n = 0,1,2,...$ A weight function that is compatible with the $y$-configuration space is $(1+y)^\alpha (1-y)^\beta$, where $\alpha$ and $\beta$ are real parameters with ranges that depend on the boundary conditions and square integrability. Our choice of polynomial is the Jacobi polynomial $P_n^{(\mu,\nu)}(y)$ because it is defined on the interval $y \in [-1,+1]$ and has the same type of weight function in their orthogonality relation [7]. The real parameters $\mu$ and $\nu$ are greater than $-1$. Thus, we write the basis functions as

$$\phi_n(y) = A_n (1+y)^\alpha (1-y)^\beta P_n^{(\mu,\nu)}(y), \tag{6}$$

where $A_n$ is a properly chosen normalization constant. We expand the wavefunction as $\psi(y,E) = \sum_{n=0}^{\infty} f_n(E)\phi_n(y)$, where $\{f_n\}$ are the (Fourier) expansion coefficients. A complete solution of the problem is obtained if all $\{f_n(E)\}_{n=0}^{\infty}$ are determined. For the present setting, it might be sufficient to state that completeness and square integrability will guarantee bounded-ness and convergence of the series. Now, it is only for a limited special set of coordinate transformation $y(\lambda x)$ that the matrix representation of the wave operator in the basis $\{\phi_n\}$ could become tridiagonal. That is, if we define the wave operator $J(y)$ as $J = (y')^2 \frac{d^2}{dy^2} + y'' \frac{d}{dy} + U(y)$, then its matrix elements $J_{nm} = \langle \phi_n | J | \phi_m \rangle = 0$ for $|n-m| \geq 2$.

Now, we are ready to consider variations on the classic problem of a particle in an infinite potential well with a flat bottom, which is usually one of the first problems to be solved exactly by students of quantum mechanics [8]. However, any deformation of the shape of the bottom of the well does not result in an exact solution. Nonetheless, we will show that by reshaping the bottom in a special way we could still obtain an exact solution using the tridiagonal representation approach. To do that, we consider the coordinate transformation $y = \cos(k\lambda x)$, where $k \in \mathbb{N} = 1, 2, 3...$, $x \in [0, L]$, and $\lambda = \pi/L$. The wave equation (5) becomes

$$\left[ (1-y^2)\frac{d^2}{dy^2} - y\frac{d}{dy} + \frac{U(y)}{k^2} \right] \psi(y, E) = 0. \tag{7}$$

The recursion relation and orthogonality formula of the Jacobi polynomials [7] show that the matrix representation for the wave operator $\langle \phi_n | J | \phi_m \rangle$ becomes tridiagonal only for three types of potential configurations. The one, which is relevant to our work, corresponds to

$$\frac{U}{k^2} = -A\frac{1-y}{1+y} - B\frac{1+y}{1-y} - Cy - D, \quad 2\alpha = \nu + \tfrac{1}{2}, \text{ and } 2\beta = \mu + \tfrac{1}{2}, \tag{8}$$

where $A$, $B$, $C$, and $D$ are real physical parameters and the basis parameters $\alpha$ and $\beta$ must satisfy $\alpha(\alpha - \tfrac{1}{2}) = A$ and $\beta(\beta - \tfrac{1}{2}) = B$. This gives the following potential function

$$\frac{2V(x)}{k^2\lambda^2} = \frac{A}{\cos^2\left(\tfrac{k}{2}\lambda x\right)} + \frac{B}{\sin^2\left(\tfrac{k}{2}\lambda x\right)} + C\cos(k\lambda x), \tag{9}$$

and parameter assignments: $D = A + B - 2E/k^2\lambda^2$, $\nu^2 = 4A + \tfrac{1}{4}$, and $\mu^2 = 4B + \tfrac{1}{4}$. Therefore, aside from the usual trigonometric Pöschl-Teller potential, Eq. (9) includes the new potential component $V(x) = \tfrac{1}{2}Ck^2\lambda^2 \cos(k\lambda x)$. Using the recursion relation and orthogonality formula of the Jacobi polynomials, we obtain the following matrix elements of the wave operator (in the atomic units $\hbar = m = 1$),

$$J_{nm} = \left[ \varepsilon - \left(n + \tfrac{\mu+\nu+1}{2}\right)^2 \right] \delta_{nm} - C\langle n | y | m \rangle, \tag{10}$$

where $\varepsilon = 2E/k^2\lambda^2$ and $\langle n | y | m \rangle$ is given as

$$\langle n|y|m\rangle = \tfrac{\nu^2-\mu^2}{(2n+\mu+\nu)(2n+\mu+\nu+2)}\delta_{n,m} + \tfrac{2}{2n+\mu+\nu}\sqrt{\tfrac{n(n+\mu)(n+\nu)(n+\mu+\nu)}{(2n+\mu+\nu-1)(2n+\mu+\nu+1)}}\delta_{n,m+1}$$
$$+ \tfrac{2}{2n+\mu+\nu+2}\sqrt{\tfrac{(n+1)(n+\mu+1)(n+\nu+1)(n+\mu+\nu+1)}{(2n+\mu+\nu+1)(2n+\mu+\nu+3)}}\delta_{n,m-1} \tag{11}$$

Thus, the only off-diagonal entries in $J_{nm}$ are due to the last term, which is proportional to the potential parameter $C$. Therefore, the usual diagonal representation constraint requires that $C = 0$ giving $E_n = \tfrac{1}{2}k^2\lambda^2\left(n+\tfrac{\mu+\nu+1}{2}\right)^2$, which is the well-known energy spectrum for the trigonometric Pöschl-Teller potential [8]. Now, we investigate the special case $A = B = 0$. That is, we look for a solution of the 1D infinite potential well problem with a bottom that has the cosine shape $\tfrac{1}{2}Ck^2\lambda^2\cos(k\lambda x)$. Obviously, the conventional diagonal representation method will not lead to an exact solution since it requires $C = 0$. The vanishing of $A$ and $B$, and the boundary conditions, which require that the wavefunction vanishes at $x = 0$ and $x = L$, dictate that $(\mu,\nu) = \left(+\tfrac{1}{2},+\tfrac{1}{2}\right)$ giving

$$\phi_n(y(x)) = \tfrac{1}{\sqrt{\pi}}\sin\left[(n+1)k\lambda x\right]. \tag{12}$$

For $k = 1$, these basis elements are just the eigenfunctions of the potential well with a flat bottom [8]. However, one must remember that an eigenfunction for the current problem is an infinite sum of these and it depends on the potential parameter $C$. Now, the tridiagonal matrix representation of the wave operator (10) simplifies to

$$J_{nm} = \left[\varepsilon-(n+1)^2\right]\delta_{nm} - \tfrac{1}{2}C\left(\delta_{n,m+1}+\delta_{n,m-1}\right). \tag{13}$$

Hence, the resulting three-term recursion relation for the expansion coefficients of the wavefunction becomes:

$$\varepsilon f_n = (n+1)^2 f_n + \tfrac{1}{2}C\left(f_{n-1}+f_{n+1}\right), \tag{14}$$

where $n = 1, 2, 3, \ldots$ The initial relation ($n = 0$) is $\varepsilon f_0 = f_0 + \tfrac{1}{2}C f_1$. The solution of this recursion relation is defined modulo an overall non-singular function of $\varepsilon$. If we call this function $\omega(\varepsilon)$, then we can write $f_n(\varepsilon) = \omega(\varepsilon)Q_n(\varepsilon)$. Choosing the standard normalization $Q_0 = 1$, determines all $\{Q_n(\varepsilon)\}_{n=0}^{\infty}$ for any given potential coupling $C$ as polynomials in $\varepsilon$. Completeness of the basis and wavefunction normalization imply that $\omega(\varepsilon) = 1/\sqrt{K(\varepsilon)}$, where $K(\varepsilon)$ is the kernel associated with $\{Q_n(\varepsilon)\}_{n=0}^{\infty}$ and defined as $K(\varepsilon) = \sum_n Q_n^2(\varepsilon)$. Now, the polynomials $Q_n(\varepsilon)$ do not belong to any of the known classic orthogonal polynomials. However, they are special case of the polynomials that we have found recently while working on the electric dipole problem [9]. We write the recursion relation (14) as the matrix eigenvalue equation $H|f\rangle = \varepsilon|f\rangle$, where $H$ is the infinite tridiagonal Hamiltonian matrix, $H_{nm} = (n+1)^2\delta_{nm} + \tfrac{1}{2}C\left(\delta_{n,m+1}+\delta_{n,m-1}\right)$. Therefore, the infinite set of eigenvalues of $H$, $\{\varepsilon_n\}_{n=0}^{\infty}$, are the only allowed energies for the system (i.e., the energy spectrum). From the theory of orthogonal polynomials, these are also the zeros of the polynomial $Q_N(\varepsilon)$ in the limit $N \to \infty$. It is obvious that this energy spectrum depends on the potential coupling parameter $C$. However, for large vibration numbers $n$, it is easy to see that $\varepsilon_n$ becomes independent of $C$ approaching the limit $(n+1)^2$, which is that of the potential well with a flat bottom. Table 1 gives few of the lowest energy levels of the spectrum for $C = 5.0$ and $N = 20$. One can easily check the rapid convergence of these values with $N$. Moreover, Fig. 1 shows the dependence of the energy spectrum on $C$.

Taking the limit $C \to 0$ results in the discrete energy spectrum $E_n = \frac{1}{2}k^2\lambda^2(n+1)^2$. This coincides with the well-known infinite square well (zero bottom) only for $k = 1$. Thus, turning off the potential $V(x) = \frac{1}{2}Ck^2\lambda^2\cos(k\lambda x)$ does not reproduce the case $V(x) = 0$ unless $k = 1$. This is a "Klauder-like phenomenon" [10] exhibited by this system for all $k \geq 2$.

Figure 2 is a plot of the lowest energy eigenfunctions [normalized by $\omega(\varepsilon_n)$] for $k = 1, 2,$ and 3. The potential strength parameter $C$ was chosen equal to 5.0. The wavefunction that corresponds to $\varepsilon_n$ is computed as $\psi_n(x) = \omega(\varepsilon_n)\sum_{m=0}^{N-1} Q_m(\varepsilon_n)\phi_m(x)$ for some large enough integer $N$. Numerically, the values obtained will become unstable if the number of terms, $N$, becomes too large; exceeding an integer that depends on $C$. For $C = 5.0$, our numerical routine produced the stable plots shown in the figure, but as $N$ is increased beyond $N = 13$ it becomes unstable. Attempting to evaluate the wavefunction at an energy not equal to one of the eigenvalues $\{\varepsilon_n\}$ will never achieve stable results. It will only produce rapidly increasing number of oscillations with large amplitudes. In fact, the sum of these oscillations for large $N$ leads to destructive interference that should result in zero net value for the wavefunction.

In conclusion, we have constructed a tridiagonal matrix representation of the wave operator in a complete square integrable basis. This made the wave equation equivalent to a three-term recursion relation for the expansion coefficients of the wavefunction. Therefore, the original problem could be solved by giving an exact solution of the recursion relation in terms of orthogonal polynomials. Consequently, a larger class of solvable potentials is obtained. The usual diagonal representation constraint results in a reduction to the standard solvable class. As an illustration, we obtained the energy spectrum and corresponding wave functions (in terms of the dipole polynomials) for the 1D infinite square potential well with sinusoidal bottom, which has no previously known exact solutions. The approach is easily extendable to higher dimensions with central as well as noncentral problems. Recently, we were able to find an exact solution for the noncentral electric dipole potential in 3D, $V(r,\theta) = -p\cos\theta/r^2$ [9], which was known not to belong to any of the established classes of exactly solvable potentials. We are currently studying a central 3D S-wave ($\ell = 0$) problem with 1/r singular potential that has a barrier. It is the product of the Hulthén potential $C(e^{\lambda r} - 1)^{-1}$ and an exponentially decaying function, $1 - 2e^{-\lambda r}$. This problem has an exact solution only in the tridiagonal representation. In the near future, we will also report on an exact solution for the 1D single-wave potential barrier $V(x) = C\tanh(\lambda x)/\cosh^2(\lambda x)$. We believe that the tridiagonal representation approach will enable us to enlarge the class of exactly solvable quantum problems in all space dimensions and should be easily extended to the relativistic wave equation.

**Acknowledgements:** This work is partially supported by King Fahd University of Petroleum and Minerals. The financial support by "Khaled Technical & Commercial Services" (KTeCS) is highly appreciated.

**Table captions**

**Table 1**: The lowest levels of the energy spectrum for a potential coupling parameter $C = 5.0$. All displayed digits are significant and correspond to a basis size $N = 20$.

**Table 1**

| $n$ | $\varepsilon_n$ |
| --- | --- |
| 0  | −0.5955395589892 |
| 1  | 4.3453451696558 |
| 2  | 9.3549646941810 |
| 3  | 16.2001100732554 |
| 4  | 25.1266923657196 |
| 5  | 36.0875520021223 |
| 6  | 49.0641568653649 |
| 7  | 64.0490437059898 |
| 8  | 81.0387114884928 |
| 9  | 100.0313345578344 |
| 10 | 121.0258834242971 |

**Figure captions**

**Fig. 1:** Dependence of the energy spectrum on the potential strength parameter, $C$.
**Fig. 2:** Plots of the lowest energy eigenfunctions $\psi_n(x)$ for several values of $k$ and $n$.

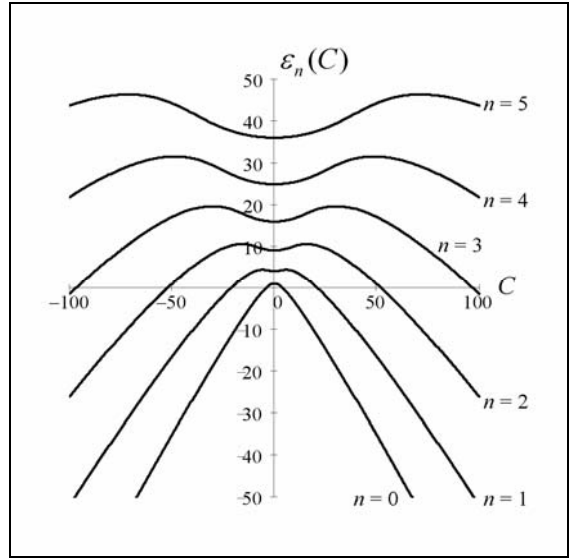

**Fig. 1**

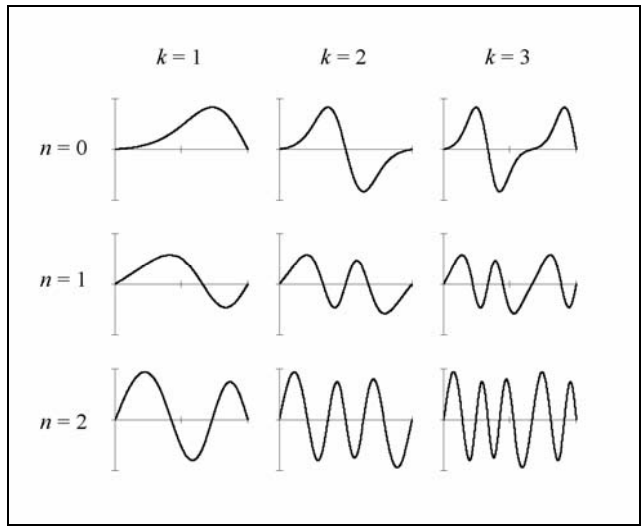

**Fig. 2**